\documentclass[10pt,conference]{IEEEtran}

\usepackage{graphicx}
\usepackage{cite}
\usepackage{amsmath}
\usepackage{amssymb}
\usepackage[dvipsnames]{xcolor}
\usepackage{algorithm}
\usepackage{algpseudocode}
\usepackage{multirow}
\usepackage{booktabs}
\usepackage{float}

\newcommand{\diag}{\mathop{\mathrm{diag}}}
\newcommand{\trace}{\mathop{\mathrm{tr}}}
\newcommand\blfootnote[1]{%
\begingroup
\renewcommand\thefootnote{}\footnote{#1}%
\addtocounter{footnote}{-1}%
\endgroup
}

\begin{document}

\title{Efficient Localization with Base Station-Integrated \\{Beyond Diagonal RIS}\vspace{-0.5em}}

\author{Mahmoud~Raeisi$^*$,
        Hui~Chen$^{\bullet}$,
        Henk~Wymeersch$^{\bullet}$,
        Ertugrul~Basar$^*$
        \\
        $^*$Koc University, Turkey,
        $^{\bullet}$Chalmers University of Technology, Sweden 
        \\
        Email: \{mraeisi19, ebasar\}@ku.edu.tr, \{hui.chen, henkw\}@chalmers.se
\vspace{-1.5em}
}

\maketitle

\begin{abstract}
This paper introduces a novel approach to efficient localization in next-generation communication systems through a base station (BS)-enabled passive beamforming utilizing beyond diagonal reconfigurable intelligent surfaces (BD-RISs). Unlike conventional diagonal RISs (D-RISs), which suffer from limited beamforming capability, a BD-RIS provides enhanced control over both phase and amplitude, significantly improving localization accuracy. By conducting a comprehensive Cramér-Rao lower bound (CRLB) analysis across various system parameters in both near-field and far-field scenarios, we establish the BD-RIS structure as a competitive alternative to traditional active antenna arrays. Our results reveal that BD-RISs achieve near active antenna arrays performance in localization precision, overcoming the limitations of D-RISs and underscoring its potential for high-accuracy positioning in future communication networks. This work envisions the use of BD-RIS for enabling passive beamforming-based localization, setting the stage for more efficient and scalable localization strategies in sixth-generation networks and beyond.
\end{abstract}


\begin{IEEEkeywords}
BD-RIS, efficient localization, passive beamforming, near-field and far-field, Cramér–Rao lower bound.
\end{IEEEkeywords}

\IEEEpeerreviewmaketitle

\vspace{-1.25em}

\section{Introduction}

\vspace{-0.5em}

In sixth-generation (6G) communication systems, high-precision localization is expected to be achieved with a single anchor, enabled by advances in millimeter wave (mmWave) communication and large antenna arrays \cite{9129075, 9145059}. 
Such technologies allow base stations (BSs) or user equipments (UEs) to measure angles of departure (AoD) and arrival (AoA) more precisely. Reconfigurable intelligent surfaces (RISs) offer a promising, cost-effective alternative to large antenna arrays by enabling passive beamforming within transceivers \cite{10515204}. For instance, the stacked intelligent metasurface (SIM) transceiver design \cite{10158690, 10515204, 10679332 , 10643881} demonstrates the potential of passive beamforming with benefits such as ultra-fast processing, reduced cost and complexity, and lower energy consumption.
Beyond diagonal RIS (BD-RIS) advances RIS technology by enabling control over both the amplitude and phase of impinging waves, offering higher flexibility in passive beamforming \cite{li2023reconfigurable, 9514409}. 
Notably, unlike active or absorptive RIS \cite{10251110}, BD-RIS controls both phase and amplitude of the impinging signal under a unitary constraint, without amplification or absorption.

The fundamental limits of mmWave multiple-input multiple-output (MIMO) systems with large RISs, focusing on channel estimation, localization, and orientation error bounds, are studied in \cite{9129075}. The study of \cite{10308579} investigates the use of a BD-RIS at the transmitter for massive MIMO, using manifold optimization to enhance spectral efficiency with fewer active antennas. Similarly, \cite{chen2024transmitter} shows that BD-RIS enhances both communication and sensing performance in mmWave integrated sensing and communication (ISAC) systems, significantly reducing power consumption and presenting a promising solution for future applications. 
RIS-aided near-field (NF) localization is further explored in \cite{10017173, 10436860}, showing the potential for NF localization by exploiting spherical wavefront. A SIM-aided ISAC system, with shared elements for communication and sensing, is proposed in \cite{10643881}, while the role of SIM for direction-of-arrival (DoA) estimation is examined in \cite{10622963, 10557708}. 

In scenarios such as those studied in \cite{10017173, 10643881, 10622963, 10557708}, where part or all of the channel operates in the NF, the signal undergoes amplitude variations across different RIS elements alongside phase changes. Diagonal RIS (D-RIS), however, only enables phase adjustment, whereas BD-RIS allows control over both amplitude and phase. Despite this added flexibility, the localization potential of BD-RIS remains unexplored. This gap in the literature motivates our investigation of BD-RIS’s potential for precision localization in such scenarios. The main contributions of this paper are as follows:

\vspace{-0.25em}

\begin{enumerate}
    \item We propose a BS-enabled passive beamforming system using a BD-RIS for downlink efficient localization, a scenario that has not been previously explored. 
    \item We analyze the beamforming gain of the proposed system, demonstrating the superiority of BD-RISs over traditional structures. Additionally, Cramér-Rao lower bound (CRLB) analysis in both NF and far-field (FF) scenarios is performed.
    \item We conduct a comprehensive analysis, evaluating localization performance across various system parameters. This detailed analysis offers new insights into the potential of BD-RIS, highlighting its capability to enhance precision localization from multiple perspectives.
\end{enumerate}

\blfootnote{
\textit{Notation}: Bold lowercase and uppercase symbols denote vectors and matrices, respectively. The notations $(.)^\mathsf{H}$, $(.)^\mathsf{T}$, $||.||$, $\diag(.)$, and $\angle$ signify the Hermitian, transpose, norm, diagonalization, and phase of a complex number. $\Re(.)$ and $\Im(.)$ refer to the real and imaginary components, while $\odot$ represents the Hadamard (element-wise) product. $\mathbb{R}$ and $\mathbb{C}$ represent the sets of real and complex numbers, respectively. The derivative of $\boldsymbol{b}$ with respect to $q$ is denoted as $\dot{\boldsymbol{b}}_q = \partial \boldsymbol{b} / \partial q$. The notation $[.]_\ell$ indicates the $\ell$-th element of a vector, while $[.]_{\ell,s}$ specifies the entry in the $\ell$-th row and $s$-th column of a matrix. $\mathcal{CN}(\mu,\sigma^2)$ represents a complex Gaussian distribution with mean $\mu$ and variance $\sigma^2$. $\mathbf{I}_n$ denotes the $n \times n$ identity matrix, and $\mathcal{U}(a,b)$ signifies a uniform distribution between $a$ and $b$. Furthermore, $[a:\Delta:b]$ represents a discrete sequence from $a$ to $b$ in steps of $\Delta$. This paper analyzes two scenarios: Scenario 1 focuses on the NF case, while Scenario 2 addresses the FF case. To maintain consistency and simplify notation, any quantity or variable specific to Scenario  $i \in \{ 1, 2 \}$ is indexed with the subscript  $i$  throughout this paper.
}

\vspace{-2em}
\section{System, Channel, and Signal Model}
\vspace{-0.5em}
This section outlines the proposed system, channel, and signal model, which leverages BS-enabled passive beamforming assisted by a BD-RIS to facilitate downlink localization.
\vspace{-0.85em}
\subsection{System Model}
\vspace{-0.45em}

\begin{figure}
    \centering
    \includegraphics[width = \columnwidth]{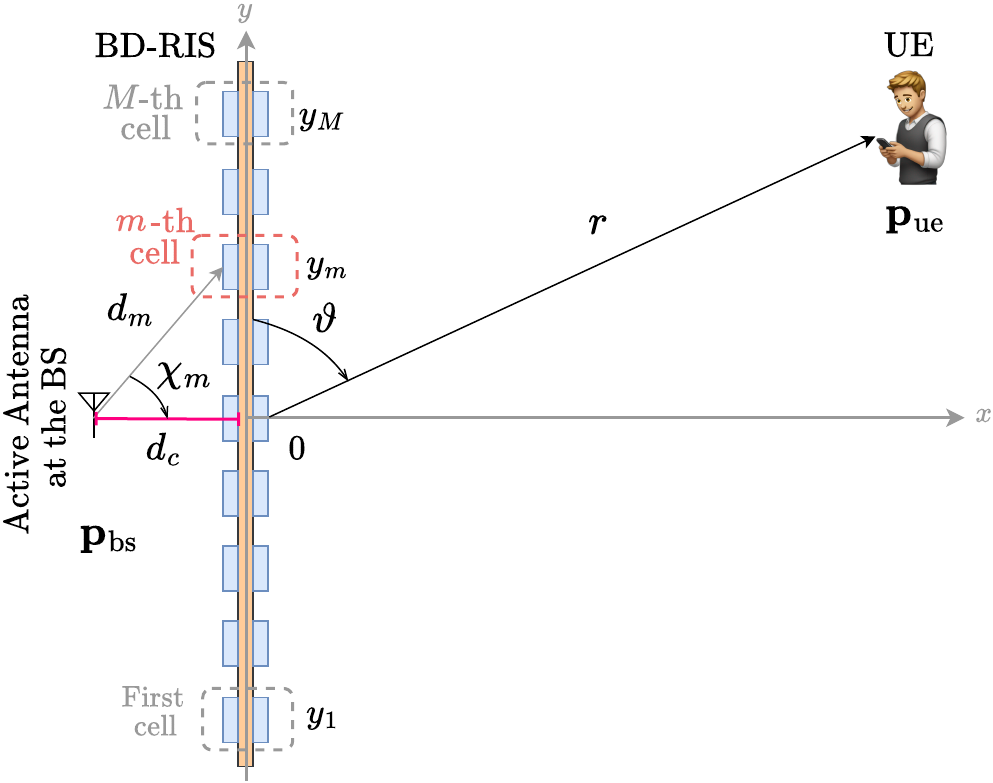}
    \caption{System model for the proposed BS-integrated BD-RIS.}
    \label{fig:System Model}
\end{figure}

Fig.~\ref{fig:System Model} illustrates the proposed system designed to enable passive beamforming at the BS in a single-input single-output (SISO) communication system, where a linear fully-connected BD-RIS is integrated with the BS to emulate a multiple-input single-output (MISO) system.\footnote{This work is the first to explore the integration of fully-connected BD-RIS at the BS for passive beamforming in localization. We focus on 2D localization to simplify the analysis and provide a clear understanding of the system’s behavior, reserving the more complex 3D analysis for future research.}  The BD-RIS consists of $M$ cells, each containing two elements: one faces the active antenna, while the other on the opposite side transmits signals toward the UE \cite{9913356}. The inter-element spacing on both sides of the BD-RIS is $\delta = \lambda/2$, where $\lambda$ denotes the wavelength of the carrier frequency. Accordingly, the array aperture of BD-RIS can be calculated as $D = (M-1)\delta$. All elements are internally connected, forming a fully-connected BD-RIS structure \cite{li2023reconfigurable}. For passive beamforming to occur at the BS, the total energy of the signal emitted by the active antenna must pass through the BD-RIS. As a result, the BD-RIS operates exclusively in transmissive mode, with its reflective mode inactive \cite{9913356}.

For simplicity, we assume that the center of the BD-RIS is located at the origin of the reference Cartesian system, i.e., $\boldsymbol{p}_{\textrm{ris}} = \left[0, 0 \right]^\mathsf{T}$, and is aligned along the $y$-axis. Therefore, the position of the $m$-th cell is given by $\boldsymbol{p}_{m} = [0,y_m]^\mathsf{T}$, where $y_m = m - \frac{M+1}{2}$. 
The BS is positioned at \(\boldsymbol{p}_{\textrm{bs}} = \left[-d_c, 0 \right]^\mathsf{T}\), where $d_c$ denotes the distance between the BS’s active antenna and the center of the BD-RIS and is on the order of a few wavelengths. The location of the UE, \(\boldsymbol{p}_{\textrm{ue}}\), is unknown and must be estimated. Accordingly, the distance between the RIS and the UE is expressed as $r = \lVert \boldsymbol{p}_{\textrm{ue}} - \boldsymbol{p}_{\textrm{ris}} \rVert$. As illustrated in Fig.~\ref{fig:System Model}, the UE is located at an angular position of $\vartheta = \arcsin{( \left([\boldsymbol{p}_{\textrm{ue}}]_1 - [\boldsymbol{p}_{\textrm{ris}}]_1) / r\right)}$ relative to the $y$-axis.

\vspace{-0.3em}

\subsection{Channel Model}
\vspace{-0.4em}
In the proposed system model, two distinct communication channels are defined: $\boldsymbol{g}[n] \in \mathbb{C}^{M \times 1}$ for the BS-RIS channel (also known as transmission coefficient vector \cite{10158690, 10515204, 10679332 , 10643881}) and $\boldsymbol{h}[n] \in \mathbb{C}^{1 \times M}$ for the RIS-UE channel. Here, $n \in [-K,K]$ is an integer representing the $n$-th subcarrier of the employed orthogonal frequency division multiplexing (OFDM) system, with a total of $N = 2K + 1$ subcarriers. 


\subsubsection{BS-RIS Channel Model}
Given the close proximity of the RIS to the BS’s active antenna, the BS-RIS channel requires a distinct model to capture its unique propagation characteristics. The channel coefficient between the active antenna and the  $m$-th RIS element at the central frequency is modeled using Rayleigh-Sommerfeld diffraction theory for NF propagation as $\left[\boldsymbol{g}[0]\right]_m = \frac{A \cos \chi_m}{d_m} \left( \frac{1}{2 \pi d_m} -  \frac{\jmath}{\lambda} \right) e^{  \jmath 2 \pi d_m / \lambda  }$ \cite{10158690, 10515204, 10679332 , 10643881}, where $A = (\lambda/2)^2$ represents the area of each passive element, $d_m$ is the distance between the active antenna and the $m$-th cell, and $\chi_m$ is the angular displacement between the line connecting the active antenna at the BS and the $m$-th cell of the BD-RIS, and the axis perpendicular to the BD-RIS surface. Accordingly, the BD-RIS channel model for the $n$-th subcarrier is as \cite{10679332}
\vspace{-0.2em}
\begin{equation}\label{eq:BS-RIS channel model}
    \left[\boldsymbol{g}[n]\right]_m = \frac{A \cos \chi_m}{d_m} \left( \frac{1}{2 \pi d_m} -  \frac{\jmath}{\lambda} \right) e^{  \jmath 2 \pi d_m \left( \frac{1}{\lambda} -  \frac{n \Delta_f}{c}\right)  },
\vspace{-0.5em}
\end{equation}
where $\Delta_f = B/N$ denotes the subcarrier spacing with $B$ is the total bandwidth, and $c$ is the speed of light.

\subsubsection{RIS-UE Channel Model}
The RIS-UE channel depends on the RIS-UE distance  $r$  and may be either NF or FF. Notably, the NF model remains valid universally, automatically simplifying to the FF model at far distances. In this paper, however, we treat them separately to allow for analysis of both models.

\vspace{-0.5em}
\begin{itemize}
    \item \textit{Scenario 1 (NF Channel Model)}: If $0.62 \sqrt{{D^3}/{\lambda}} < r < {2D^2}/{\lambda}$, the UE is located in the Fresnel (radiative) NF region of the RIS \cite{10436860, 10017173}. In this case, the RIS-UE channel is modeled as a line-of-sight (LoS) narrowband channel ($n=0$ and $N=1$), as follows \cite{10135096}:\footnote{To improve readability, we omit the subcarrier index when exclusively discussing Scenario 1 throughout this paper (i.e., $\boldsymbol{h}_1[0]$ becomes $\boldsymbol{h}_1$).} 
    \vspace{-0.5em}
    \begin{equation}\label{eq:NF channel}
        \boldsymbol{h}_1[0] = \boldsymbol{h}_1 = \beta_1 \boldsymbol{a}_1^\mathsf{H}(r,\vartheta),
        \vspace{-0.5em}
    \end{equation}
    where $\beta_1 = \frac{\lambda}{4 \pi r} e^{  - \jmath \frac{2 \pi}{\lambda}r }$ represents the complex channel gain\cite{10436860}. The term $\boldsymbol{a}_1(r, \vartheta) \in \mathbb{C}^{M \times 1}$ denotes the NF array response vector at the RIS, defined as \cite{10135096}:
    \vspace{-0.5em}
    \begin{equation}\label{eq: NF steering vector}
        \left[ \boldsymbol{a}_1(r,\vartheta) \right]_m = \frac{1}{\sqrt{M}} e^{ - \jmath \frac{2\pi}{\lambda} (r_m(r,\vartheta) - r) },
        \vspace{-0.5em}
    \end{equation}
    where $r_m (r, \vartheta) = \sqrt{r^2 + y_m^2 \delta^2 - 2ry_m\delta \cos(\vartheta)}$ is the distance between the UE and the $m$-th cell.

    \item \textit{Scenario 2 (FF Channel Model)}: If $r > {2D^2}/{\lambda}$, the UE is located in the FF region \cite{9145059}. In this case, we model the RIS-UE channel as a LoS wideband channel. The channel for the $n$-th subcarrier is expressed as follows \cite{9129075, 9124848}:
    \vspace{-0.5em}
    \begin{equation}\label{eq:FF channel}
        \boldsymbol{h}_2 [n] = \beta_2 e^{ -\jmath 2 \pi \tau n \Delta_f }\boldsymbol{a}_2^\mathsf{H}(\vartheta),
        \vspace{-0.5em}
    \end{equation}
    where $\beta_2 = \frac{\lambda}{4 \pi r} e^{ \jmath \varphi }$ represents the complex channel gain, with $\varphi \sim \mathcal{U}(0, 2\pi)$ \cite{9500281}. Here, $\tau =  r/c$ is time of arrival (ToA) for RIS-UE channel,\footnote{To simplify analysis, we assume tight synchronization between the BS and UE, leaving asynchronous scenario for the future work.} and $\boldsymbol{a}_2(\vartheta) \in \mathbb{C}^{M \times 1}$ is the array response (steering) vector at the RIS, defined as follows \cite{9124848, 9129075, 10188340}:
    \vspace{-0.5em}
    \begin{equation}\label{eq: FF steering vector}
        [\boldsymbol{a}_2(\vartheta)]_m = \frac{1}{\sqrt{M}} e^{ \jmath 2\pi (m-1) \frac{\delta}{\lambda} \cos{\vartheta} }.
        \vspace{-0.5em}
    \end{equation}
\end{itemize}

\vspace{-1.5em}

\subsection{Signal Model}

To facilitate downlink localization, the BS systematically sweeps the environment, transmitting $T_i$ pilot signals to various zones within the area of interest over sequential time slots in the $i$-th scenario. Without loss of generality, we assume a unit transmit pilot symbol. Thus, the received signal at the UE for the $n$-th subcarrier in Scenario $i$ during the $t$-th time slot is as
\vspace{-0.5em}
\begin{equation}\label{eq: Signal Model}
    y_{i,t}[n] = \sqrt{P} \boldsymbol{h}_i[n] \boldsymbol{\Omega}_{i,t} \boldsymbol{g}[n] + w_{i,t}[n],
    \vspace{-0.5em}
\end{equation}
where $P$ is the transmitted power, and $w_{i,t} \sim \mathcal{CN}(0,\sigma^2)$ represents the additive noise component during the $t$-th time slot in the $i$-th scenario.
Here, $\boldsymbol{\Omega}_{i,t} \in \mathbb{C}^{M \times M}$ is the RIS phase shift matrix during $t$-th time slot in the $i$-th scenario. To ensure that the BD-RIS is power-lossless, its phase shift matrix should satisfy unitary constraint, i.e., $\boldsymbol{\Omega}_{i,t}^\mathsf{H} \boldsymbol{\Omega}_{i,t} = \boldsymbol{I}_M$. Additionally, the phase shift matrix must be symmetric, i.e., $\boldsymbol{\Omega}_{i,t} = \boldsymbol{\Omega}_{i,t}^\mathsf{T}$, ensuring identical phase adjustments between each pair of passive elements, allowing a single phase shifter per element pair. 
Ultimately, for the purpose of the CRLB analysis, we define the noiseless component of the received signal as
\vspace{-0.5em}
\begin{equation}
    \mu_{i,t}[n] = \boldsymbol{h}_i[n] \boldsymbol{\Omega}_{i,t} \boldsymbol{g}[n] = \boldsymbol{h}_i[n] \boldsymbol{\zeta}_{i,t}[n],
    \vspace{-0.35em}
\end{equation}
where $\boldsymbol{\zeta}_{i,t}[n] = \boldsymbol{\Omega}_{i,t} \boldsymbol{g}[n]$.


\subsection{Pre-defined Codebook for BD-RIS Configuration}
In each scenario, a pre-defined codebook enables systematic environmental sweeping, with different codewords selected in each time slot to ensure full coverage. Since $T_i$ symbols are required, the codebook must contain $T_i$ codewords, each satisfying the unitary and symmetry constraints outlined above. Takagi’s decomposition, as described in \cite{10187688}, is used to construct this codebook, with detailed steps provided in Algorithm \ref{Alg: Takagi's Decomposition}.
\vspace{-0.5em}
\begin{algorithm}
\footnotesize
\caption{Pre-defined codebook for BD-RIS configuration}
\label{Alg: Takagi's Decomposition}
\begin{algorithmic}[1] 
    \State Define $T_i$ uniformly distributed sweeping points/angles for NF/FF:
    \begin{enumerate}
        \item[(NF)] Define $T_1$ sweeping points as $\{\boldsymbol{s}_1, \boldsymbol{s}_2, \dots, \boldsymbol{s}_{T_1}\}$ with \{$\boldsymbol{s}_t = [r_t, \vartheta_t]^\mathsf{T} | r_t \in [ \varrho_{\textrm{min}}:\Delta_r:\varrho_{\textrm{max}}]$, $\vartheta_t \in [0: \Delta_{\vartheta}  :180^\circ]$\}, where $\varrho_{\textrm{min}} > 0.62\sqrt{\frac{D^3}{\lambda}}$ and $\varrho_{\textrm{\textrm{max}}}<\frac{2 D^2}{\lambda}$.
        \item[(FF)] Define $T_2$ sweeping angles as $\{\vartheta_1, \vartheta_2, \dots, \vartheta_{T_2}\}$ with $\vartheta_t \in [0: \Delta_{\vartheta} :180^\circ]$.
    \end{enumerate}
    \State Compute the transmit steering vector at the transmit terminal of BD-RIS for the $t$-th codeword (associated with the $t$-th time slot):
    \begin{enumerate}
        \item[(NF)] Compute toward location $\boldsymbol{s}_t$ using (\ref{eq: NF steering vector}).
        \item[(FF)] Compute toward direction $\vartheta_t$ using (\ref{eq: FF steering vector}).
    \end{enumerate}
    \State Calculate $\boldsymbol{u}_T = \boldsymbol{g}[0] / \lVert \boldsymbol{g}[0] \rVert$; set $\boldsymbol{u}_{R} = \boldsymbol{a}_1(r_t, \vartheta_t)$ for NF or $\boldsymbol{u}_{R} = \boldsymbol{a}_2( \vartheta_t)$ for FF.
    \State Calculate the symmetric matrix $\boldsymbol{A}$ and decompose it using standard singular value decomposition (SVD) as $\boldsymbol{A} = \boldsymbol{u}_{R} \boldsymbol{u}_T^\mathsf{H} + (\boldsymbol{u}_{R} \boldsymbol{u}_T^\mathsf{H})^\mathsf{T} = \boldsymbol{U} \boldsymbol{\Sigma} \boldsymbol{V}^\mathsf{H}$.
    \State Compute $\boldsymbol{\nu} = \text{diag}(\boldsymbol{U}^\mathsf{H} \boldsymbol{V}^*)$, followed by $\boldsymbol{\phi} = \angle \boldsymbol{\nu} / 2$. 
    \State After computing $\boldsymbol{Q} = \boldsymbol{U} \, \text{diag}(\exp(\jmath \boldsymbol{\phi}))$, Takagi’s decomposition can be expressed as $\boldsymbol{A} = \boldsymbol{Q} \boldsymbol{\Sigma} \boldsymbol{Q}^\mathsf{T}$; thus, the $t$-th codeword for the $i$-th scenario is obtained as $\boldsymbol{\Omega}_{i,t} = \boldsymbol{Q}\boldsymbol{Q}^\mathsf{T}$.
\end{algorithmic}
\end{algorithm}

\vspace{-1em}

\section{Fisher Information Analysis}

In this section, we calculate the position error bound (PEB), which represents a theoretical lower bound on the achievable accuracy of position estimation. The positional parameters for scenario $i$ are expressed as $\boldsymbol{\xi}_{po,i} = [\boldsymbol{p}_{\textrm{ue},i}^\mathsf{T}, \Re{(\beta_i)}, \Im{(\beta_i)}]^\mathsf{T}$. However, before proceeding, we must first derive the CRLB for the channel parameters. In the $i$-th scenario, the Fisher information matrix (FIM), $\boldsymbol{\mathcal{F}}_{ch,i} \in \mathbb{R}^{4 \times 4}$, with respect to the channel parameters set $\boldsymbol{\xi}_{ch,i}$ is obtained using the Slepian-Bangs formula \cite{kay1993fundamentals} as follows:
\vspace{-0.5em}
\begin{equation}\label{eq: FIM of channel parameters}
    \boldsymbol{\mathcal{F}}_{ch,i} = \frac{2P}{\sigma^2} \sum_{t = 1}^{T_i} \sum_{n = -K}^{K} \Re \left\{ \left( \frac{\partial \mu_{i,t}[n]}{\partial \boldsymbol{\xi}_{ch,i}} \right) \left( \frac{\partial \mu_{i,t}[n]}{\partial \boldsymbol{\xi}_{ch,i}} \right) ^\mathsf{H} \right\}.
\end{equation}
The channel parameter vector for NF and FF scenarios are defined as $\boldsymbol{\xi}_{ch,1} = [r, \vartheta, \Re{(\beta_1)}, \Im{(\beta_1)}]^\mathsf{T}$ and $\boldsymbol{\xi}_{ch,2} = [\tau, \vartheta, \Re{(\beta_2)}, \Im{(\beta_2)}]^\mathsf{T}$, respectively.
The mathematical derivations for obtaining $\boldsymbol{\mathcal{F}}_{ch,i}$ are provided in Appendix \ref{apndx: FIM of channel parameters}. Consequently, the CRLB for $[\boldsymbol{\xi}_{ch,i}]_\ell$ is calculated as:
\vspace{-0.5em}
\begin{equation}
    \eta_\ell = \textrm{CRLB}\left([\boldsymbol{\xi}_{ch,i}]_\ell\right) =  \sqrt{[\boldsymbol{\mathcal{F}}_{ch,i}^{-1}]_{\ell,\ell}}.
    \vspace{-0.5em}
\end{equation}
\begin{table}[t]
\scriptsize
  \centering
  \caption{Computer Simulation parameters.}
  \vspace{-1em}
  \begin{tabular}{c c c c}
  \toprule
  \textbf{Parameter} & \textbf{Value} & \textbf{Parameter} & \textbf{Value} \\
  \toprule
  $\boldsymbol{p}_{\textrm{ris}}$ & $[0, 0]^\mathsf{T}$ & $T_1$ & $500$ \\
  $\boldsymbol{p}_{\textrm{ue},1}$ & $[12, 8]^\mathsf{T}$ m & $T_2$ & $100$ \\
  $\boldsymbol{p}_{\textrm{ue},2}$ & $[60, 40]^\mathsf{T}$ m & $\Delta_r$ & $10$ m \\
  $d_c$ & $0.5\lambda$ \cite{10515204,10158690} & $\Delta_{\vartheta}$ & $1.8^{\circ}$ \\
  $M$ & $101$ & $\varrho_{\textrm{min}}$ & $5$ m \\
  $P$ & $20$ dBm & $\varrho_{\textrm{max}}$ & $45$ m \\
  $c$ & $3 \times 10^8$ m/s & $N_1$ & $1$ \\
  $f_c = c/\lambda$ & $28$ GHz & $N_2$ & $501$ \\
  $\Delta_f$ & $120$ KHz & $B$ & $N_i \Delta_f$ \\
  $\delta$ & $\lambda/2$ m & $\sigma^2$ & $-174 + 10\log(B)$ dBm \\
  \bottomrule
  \end{tabular}
  \label{tab:System parameters}
\end{table}
Once $\boldsymbol{\mathcal{F}}_{ch,i}$ is calculated, the FIM of the positional parameters can be obtained as $\boldsymbol{\mathcal{F}}_{po,i} = \boldsymbol{\mathcal{J}}_i^\mathsf{T} \boldsymbol{\mathcal{F}}_{ch,i} \boldsymbol{\mathcal{J}}_i$, where $\boldsymbol{\mathcal{J}}_i \in \mathbb{R}^{4 \times 4}$ is the Jacobian matrix, defined as $\left[ \boldsymbol{\mathcal{J}}_i \right]_{\ell,s} = {\partial \left[ \boldsymbol{\xi}_{ch,i} \right]_{\ell}}/{\partial \left[ \boldsymbol{\xi}_{po,i} \right]_s}.$
The mathematical derivation of $\boldsymbol{\mathcal{J}}_i$ is provided in Appendix \ref{apndx: Jacobian Matrix}. Ultimately, the $\textrm{PEB}$ for Scenario $i$ is calculated as follows:
\vspace{-0.5em}
\begin{equation}
    \textrm{PEB}_i = \sqrt{\trace{\left( [\boldsymbol{\mathcal{F}}_{po,i}^{-1}]_{1:2,1:2} \right)}}.
    \vspace{-0.5em}
\end{equation}

\vspace{-0.15em}

\section{Simulation Results and Analytical Insights}

\vspace{-0.35em}

This section presents the CRLB analysis of the proposed architecture. The assumed values for various parameters are listed in Table \ref{tab:System parameters}, unless stated otherwise.
For comparison, we evaluate two different benchmarks: 
\begin{itemize}
    \item \textit{Benchmark 1 (D-RIS)}: A cell-wise single-connected BD-RIS operating in transmissive mode. It is also called simultaneously transmitting and reflecting RIS (STAR-RIS) \cite{9913356}. When integrated within a BS, it functions as a single-layer stacked intelligent metasurface \cite{10158690, 10515204, 10679332 , 10643881, 10622963, 10557708, 9913356}. Due to its single-connected structure and diagonal phase shift matrix, which aligns with the characteristics of traditional D-RIS, we refer to it as D-RIS in this paper for clarity and to highlight its diagonal structure.

    \item \textit{Benchmark 2 (AAA)}: The BS equipped with a linear active antenna array (AAA) with $M$ antenna elements.
\end{itemize}

In each benchmark, a pre-defined codebook is constructed for beam-sweeping to facilitate UE localization. For both benchmarks, steps $1-3$ in Algorithm \ref{Alg: Takagi's Decomposition} are first performed. In Benchmark 1, the $t$-th codeword is generated as $\boldsymbol{\Omega}_{i,t} = \diag\left(e^{-\jmath \arg(\boldsymbol{u}_T \odot \boldsymbol{u}_R^*)}\right)$ \cite{10187688}, while $\boldsymbol{u}_R$ is used to create the $t$-th codeword for Benchmark 2.

\begin{figure}
    \centering
    \includegraphics[width = \columnwidth]{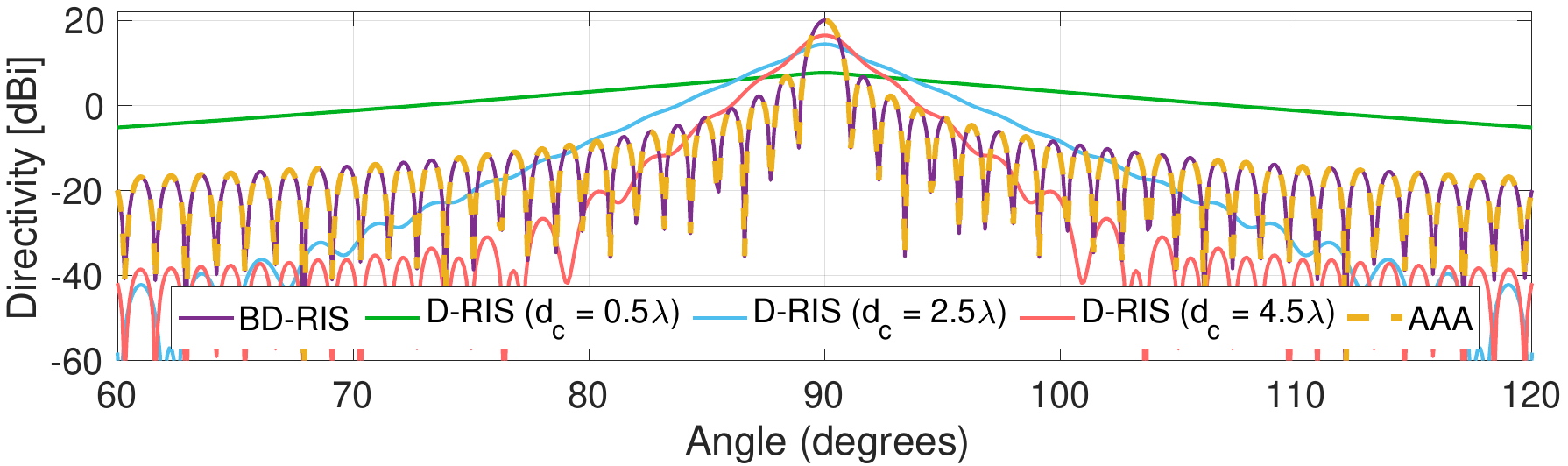}
    \caption{Transmit beamforming patterns for BD-RIS and benchmark configurations, calculated using the effective passive beamforming vector $\boldsymbol{\zeta} = \boldsymbol{\Omega} \boldsymbol{g}[0]$.}
    \label{fig:BeamPattern}
\end{figure}

\begin{figure}
    \centering
    \includegraphics[scale = 0.34]{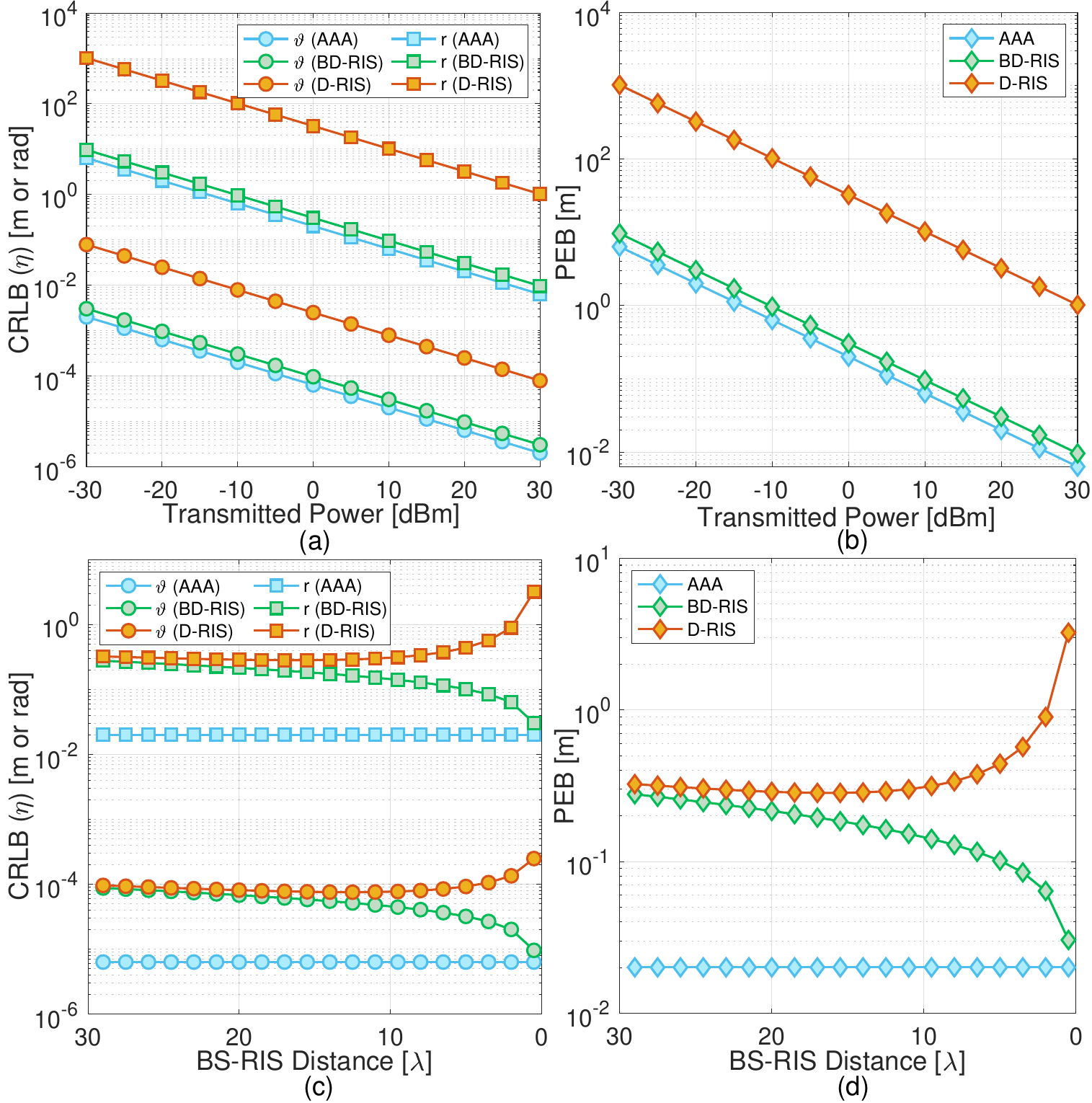}
    \caption{CRLB and PEB analysis for Scenario 1 (NF): (a) CRLB for $r$ and $\vartheta$ vs. transmitted power, (b) PEB vs. transmitted power, (c) CRLB for $r$ and $\vartheta$ vs. BS-RIS distance, and (d) PEB vs. BS-RIS distance.}
    \label{fig:NF CRLB analysis}
\end{figure}

\vspace{-0.7em}

\subsection{Beamforming Performance Analysis}

\vspace{-0.15em}

Placing the RIS closer to the BS reduces the multiplicative path loss in the overall BS-UE channel \cite{raeisi2024comprehensive}. However, as shown in Fig.~\ref{fig:BeamPattern}, D-RIS exhibits reduced beamforming performance as $d_c$ decreases. This decline occurs because the BS-RIS channel vector $\boldsymbol{g}$ shows greater variations in both magnitude and phase with a shorter $d_c$. Since D-RIS can only adjust the phase, it cannot compensate for these magnitude variations, thereby limiting its beamforming capability \cite{9514409}.
In contrast, the fully-connected BD-RIS can adjust both phase and amplitude of impinging waves, allowing it to sustain high beamforming gain even at close proximity to the BS. As shown in Fig.~\ref{fig:BeamPattern}, this capability enables BD-RIS to achieve beamforming performance comparable to the AAA structure. Although BD-RIS matches AAA in beamforming gain, AAA is still expected to outperform it due to reduced path loss, as it lacks the cascaded channels effect. Next, we evaluate BD-RIS performance against the benchmarks in terms of PEB and CRLB on channel parameters.

\begin{figure}
    \centering
    \includegraphics[width = \columnwidth]{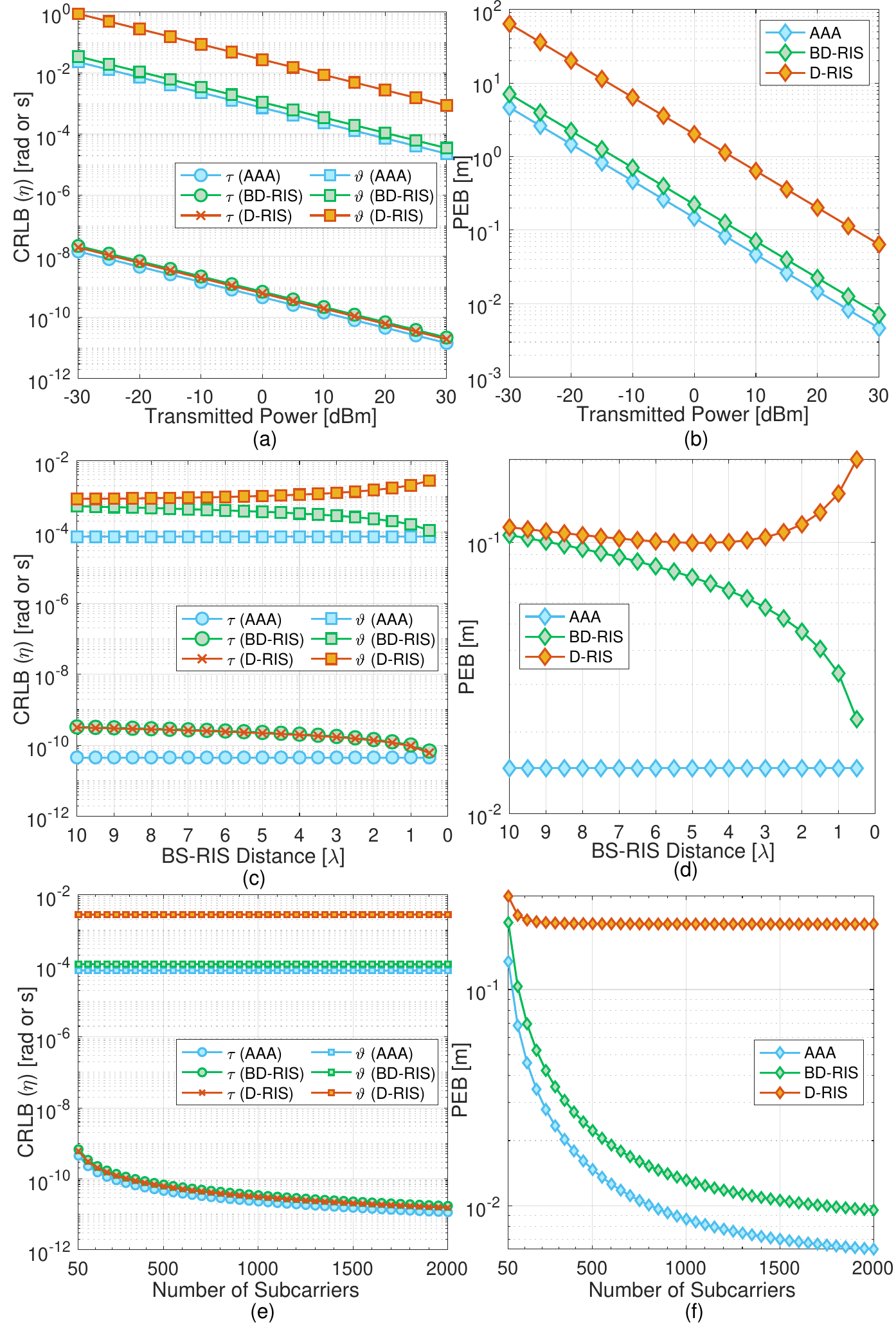}
    \caption{CRLB and PEB analysis in Scenario 2 (FF): (a) CRLB for $\tau$ and $\vartheta$ vs. transmitted power, (b) PEB vs. transmitted power, (c) CRLB for $\tau$ and $\vartheta$ vs. BS-RIS distance, (d) PEB vs. BS-RIS distance, (e) CRLB for $\tau$ and $\vartheta$ vs. number of subcarriers, and (f) PEB vs. number of subcarriers.}
    \label{fig:FF CRLB analysis}
\end{figure}

\vspace{-0.6em}

\subsection{CRLB Analysis for the NF Scenario}
\vspace{-0.45em}

The CRLB analysis for Scenario 1 (NF) is shown in Fig.~\ref{fig:NF CRLB analysis}. Figs.~\ref{fig:NF CRLB analysis}(a) and (b) present the CRLB for $r$, $\vartheta$, and the PEB across different transmitted power levels, respectively. While a significant performance gap exists between D-RIS and AAA, BD-RIS achieves performance close to AAA due to similar beamforming capabilities. 
Figs.~\ref{fig:NF CRLB analysis}(c) and (d) show the CRLB for channel parameters and PEB as a function of BS-RIS distance ($d_c$), respectively. When the RIS is positioned far from the BS, BD-RIS and D-RIS exhibit similar CRLB performance. This similarity arises from the comparable beamforming gains of both structures at larger $d_c$, as illustrated in Fig.~\ref{fig:BeamPattern}. As $d_c$ decreases, BD-RIS performance improves due to reduced multiplicative path loss, with its fully-connected structure maintaining beamforming gain. In contrast, D-RIS initially benefits from reduced $d_c$, but beyond a critical threshold, the loss in beamforming gain outweighs path loss reduction, leading to a substantial CRLB and PEB performance drop.

\vspace{-0.5em}
\subsection{CRLB Analysis for the FF Scenario}

\vspace{-0.5em}

Fig.~\ref{fig:FF CRLB analysis} presents the CRLB analysis for Scenario 2 (FF), comparing BD-RIS, D-RIS, and AAA across various parameters. In Fig.~\ref{fig:FF CRLB analysis}(a), D-RIS significantly underperforms in $\vartheta$ estimation compared to BD-RIS and AAA, which achieve higher angular resolution due to their sharp beams. 
Conversely, D-RIS achieves $\tau$ estimation comparable to BD-RIS and AAA due to its broad beam pattern, which disperses energy over a wide range. This wide dispersion allows the UE to receive pilot signals even when the BS beam is not precisely aimed at it. Notably, $\tau$ estimation primarily relies on the frequency diversity of the wideband channel rather than a sharp, narrow beam, making D-RIS’s broad coverage beneficial for this purpose. This advantage is also reflected in the CRLB trends for $\tau$ in Figs.~\ref{fig:FF CRLB analysis}(c) and (e). 
Despite this advantage in $\tau$, D-RIS’s weaker angular estimation negatively impacts its PEB, while BD-RIS attains performance close to AAA, thanks to its flexibility in beam design, as shown in Fig.~\ref{fig:FF CRLB analysis}(b). 

Figs.~\ref{fig:FF CRLB analysis}(c) and (d) show the CRLB performance of the channel parameters and the PEB performance as a function of BS-RIS distance ($d_c$), respectively. The trends for $\vartheta$ and PEB closely mirror those in Scenario 1 (NF) (see Figs.~\ref{fig:NF CRLB analysis}(c) and (d)). As $d_c$ decreases, the CRLB on $\tau$ improves for both D-RIS and BD-RIS, driven by reduced multiplicative path loss and enhanced SNR, underscoring SNR’s importance in $\tau$ estimation. Note that, AAA’s performance remains stable, unaffected by $d_c$, due to its independent MISO structure.

\begin{figure}
    \centering
    \includegraphics[width=\columnwidth]{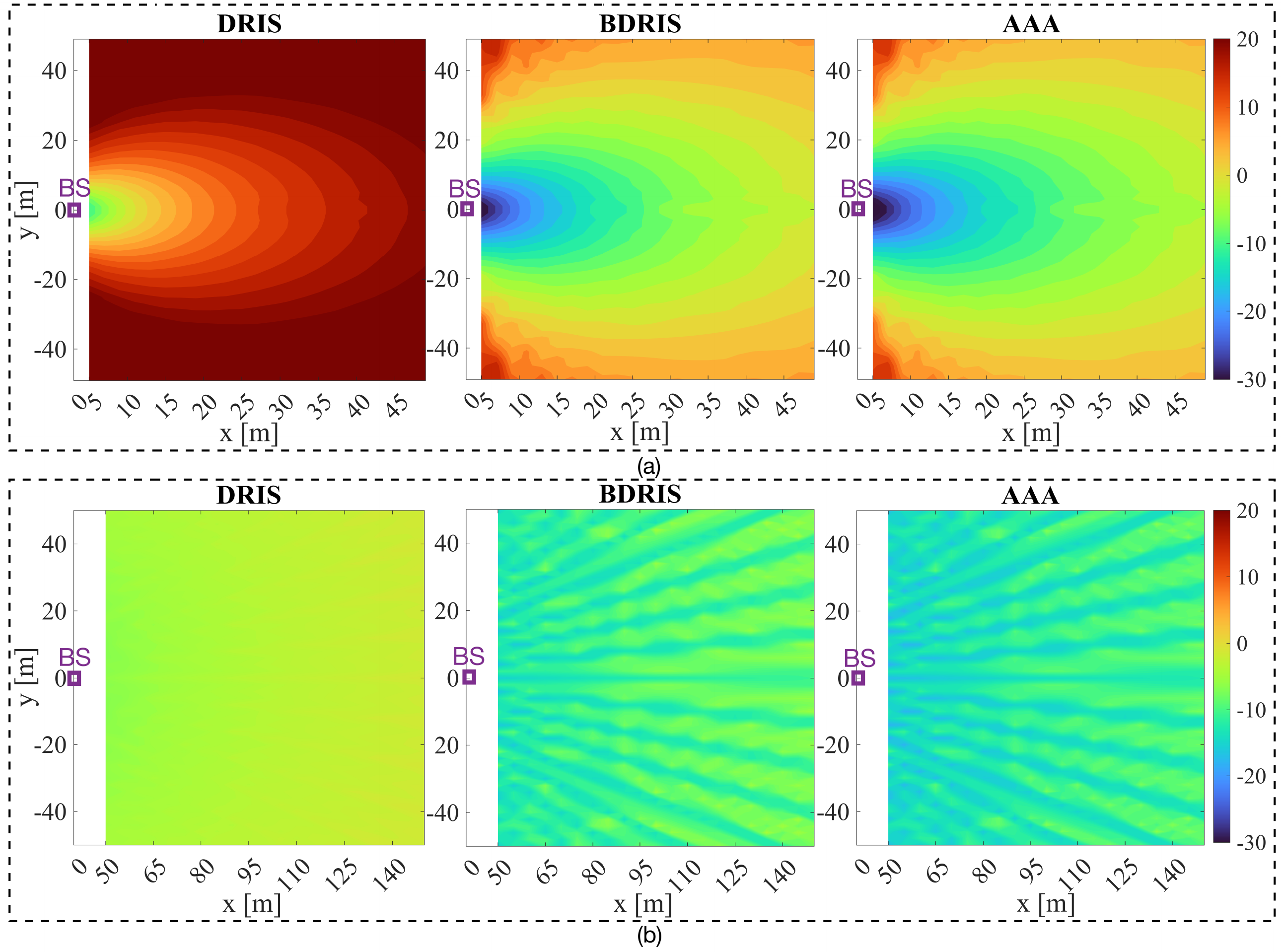}
    \caption{$10\log_{10}(\textrm{PEB})$ for different UE positions. (a) Scenario 1 (NF); (b) Scenario 2 (FF).}
    \label{fig: Different UE Positions}
\end{figure}

Figs.~\ref{fig:FF CRLB analysis}(e) and (f) respectively show the CRLB performance for channel parameters and PEB as the number of subcarriers increases. In Fig.~\ref{fig:FF CRLB analysis}(e), $\vartheta$ estimation remains unaffected by the number of subcarriers, as it depends primarily on beamforming gain. Conversely, $\tau$ estimation benefits from the frequency diversity of OFDM, showing improved CRLB performance across all setups as the number of subcarriers increases. In Fig.~\ref{fig:FF CRLB analysis}(f), D-RIS’s PEB performance plateaus after only a few subcarriers, while BD-RIS and AAA continue to improve significantly before reaching their own performance ceilings at much higher subcarrier counts. This early ceiling for D-RIS is due to its limited beamforming capability and correspondingly poorer $\vartheta$ estimation. In contrast, the lower CRLB on $\vartheta$ for BD-RIS and AAA indicates a greater capacity to leverage frequency diversity across a larger number of subcarriers, thereby enhancing PEB.

\vspace{-0.75em}

\subsection{PEB Visualization for the Area of Interest}

\vspace{-0.2em}

Figs.~\ref{fig: Different UE Positions}(a) and (b) show the PEB on a logarithmic scale for various UE positions in NF and FF scenarios, respectively. In the NF scenario, the same RIS configuration listed in Table \ref{tab:System parameters} is applied. BD-RIS achieves performance comparable to AAA, while D-RIS provides sub-meter level accuracy only for UEs within a few meters of the RIS. For the FF scenario illustrated in Fig.~\ref{fig: Different UE Positions}(b), we assume $\Delta_{\vartheta} = 5^\circ$ (instead of $\Delta_\vartheta = 1.8^\circ$, for better visualization of the beams). Similar to NF scenarios, D-RIS shows relatively poor performance, whereas BD-RIS's performance remains close to AAA. Furthermore, Fig.~\ref{fig: Different UE Positions}(b) illustrates enhanced PEB along paths where the sweeping beam aligns with the positions of UEs in both BD-RIS and AAA configurations, underscoring the critical role of high-resolution codebooks in achieving precise localization. While a direct comparison between NF (Fig.~\ref{fig: Different UE Positions}(a)) and FF (Fig.~\ref{fig: Different UE Positions}(b)) is challenging due to the higher number of samples in FF from the OFDM structure, it is evident that BD-RIS provides a more substantial improvement over D-RIS in NF than in FF. In NF scenarios, where beamforming gain is critical for estimating positional parameters $r$ and $\vartheta$, BD-RIS significantly reduces the PEB compared to D-RIS. In contrast, this enhancement is less pronounced in FF, where ToA estimation benefits more from frequency diversity rather than beamforming precision.

\vspace{-0.35em}

\section{Conclusion}
\vspace{-0.5em}
This paper has proposed a novel BS architecture integrating fully-connected BD-RIS to enable enhanced passive beamforming for efficient localization. This design offers a promising alternative to existing benchmarks, achieving localization accuracy comparable to active analog beamforming with traditional antenna arrays. Through comprehensive CRLB analysis in both NF and FF scenarios, we have demonstrated that BD-RIS offers substantial improvements over traditional D-RIS. Notably, BD-RIS shows comparable performance to AAA, establishing it as a viable solution for high-accuracy positioning in 6G networks and beyond. Our findings underline the potential of BD-RIS to bridge the gap between active and passive localization methods, paving the way for efficient, scalable localization solutions in future networks.

\vspace{-0.5em}
\appendices
\section{Detailed Derivations of $\boldsymbol{\mathcal{F}}_{ch,i}$}\label{apndx: FIM of channel parameters}

\vspace{-0.5em}

This section presents the detailed calculation of $\boldsymbol{\mathcal{F}}_{ch,i}$ by evaluating $\frac{\partial \mu_{i,t}[n]}{\partial [\boldsymbol{\xi}_{ch,i}]_{\ell}}$. Consequently, $\boldsymbol{\mathcal{F}}_{ch,i}$ is obtained using (\ref{eq: FIM of channel parameters}).

\textit{Scenario 1:} The calculations for Scenario 1 are as follows:
\begin{equation}
    [\frac{\partial \mu_{1,t}}{\partial r}, \frac{\partial \mu_{1,t}}{\partial \vartheta}]^\mathsf{T} = \beta_1 [\boldsymbol{\dot{a}}_{1,r}(r,\vartheta) , \boldsymbol{\dot{a}}_{1,\vartheta}(r,\vartheta)]^\mathsf{H} \boldsymbol{\zeta}_{1,t}[0],
\end{equation}
\begin{equation}
    [\frac{\partial \mu_{1,t}}{\partial \Re{(\beta_1)}}, \frac{\partial \mu_{1,t}}{\partial \Im{(\beta_1)}}] = [1, \jmath]\boldsymbol{a}_1^\mathsf{H}(r,\vartheta) \boldsymbol{\zeta}_{1,t}[0],
\end{equation}
where 
\begin{equation}
    [\boldsymbol{\dot{a}}_{1,r}(r,\vartheta),\boldsymbol{\dot{a}}_{1,\vartheta}(r,\vartheta)] = [{\boldsymbol{d}}_r, {\boldsymbol{d}}_{\vartheta}] \odot [\boldsymbol{a}_1(r,\vartheta), \boldsymbol{a}_1(r,\vartheta)],
\end{equation}
with 
\begin{equation}
    [{\boldsymbol{d}}_r]_m = - \jmath \frac{2\pi}{\lambda} \left( \frac{r - y_m \delta \cos{\vartheta}}{\sqrt{r^2 + y_m^2 \delta^2 - 2 r y_m \delta \cos{\vartheta}}} - 1 \right),
\end{equation}
\begin{equation}
    [{\boldsymbol{d}}_{\vartheta}]_m = - \jmath \frac{2\pi}{\lambda}  \frac{r y_m \delta \sin{\vartheta}}{\sqrt{r^2 + y_m^2 \delta^2 - 2 r y_m \delta \cos{\vartheta}}}.
\end{equation}

\textit{Scenario 2:} The calculations in this case are as follows:
\begin{equation}
    \frac{\partial \mu_{2,t} [n]}{\partial \tau} = - \jmath 2 \pi n \Delta_f \beta_2 e^{-\jmath 2 \pi \tau n \Delta_f } \boldsymbol{a}_2^\mathsf{H}(\vartheta) \boldsymbol{\zeta}_{2,t}[n], 
\end{equation}
\begin{equation}
    \frac{\partial \mu_{2,t} [n]}{\partial \vartheta} = \beta_2 e^{ -\jmath 2 \pi \tau n \Delta_f } \boldsymbol{\dot{a}}_{2,\vartheta}^\mathsf{H}(\vartheta) \boldsymbol{\zeta}_{2,t}[n],
\end{equation}
\begin{equation}
    [\frac{\partial \mu_{2,t} [n]}{\partial \Re{(\beta_2)}}, \frac{\partial \mu_{2,t} [n]}{\partial \Im{(\beta_2)}}] = [1,\jmath] e^{-\jmath 2 \pi \tau n \Delta_f } \boldsymbol{a}_2^\mathsf{H}(\vartheta) \boldsymbol{\zeta}_{2,t}[n],
\end{equation}
where $\boldsymbol{\dot{a}}_{2,\vartheta}(\vartheta) = -\jmath 2 \pi \frac{\delta}{\lambda} \cos{\vartheta} \diag(0,\dots,M-1) \boldsymbol{a}_2(\vartheta)$.


\section{Detailed Derivations of Jacobian Matrix $\boldsymbol{\mathcal{J}}_i$}\label{apndx: Jacobian Matrix}


Each channel parameter in $\boldsymbol{\xi}_{ch,i}$ must be expressed as a function of the positional parameters $\boldsymbol{\xi}_{po,i}$, after which the derivatives are readily computed.

\textit{Scenario 1:} The derivatives are as follows:
\begin{equation}
    \frac{\partial r}{\partial [\boldsymbol{p}_{\textrm{ue}}]_1} = \frac{[\boldsymbol{p}_{\textrm{ue}}]_1 - [\boldsymbol{p}_{\textrm{ris}}]_1}{\lVert \boldsymbol{p}_{\textrm{ue}} - \boldsymbol{p}_{\textrm{ris}} \rVert}, \frac{\partial r}{\partial [\boldsymbol{p}_{\textrm{ue}}]_2} = \frac{[\boldsymbol{p}_{\textrm{ue}}]_2 - [\boldsymbol{p}_{\textrm{ris}}]_2}{\lVert \boldsymbol{p}_{\textrm{ue}} - \boldsymbol{p}_{\textrm{ris}} \rVert}
\end{equation}
\begin{equation}
    \frac{\partial \vartheta}{\partial [\boldsymbol{p}_{\textrm{ue}}]_1} = - \frac{[\boldsymbol{p}_{\textrm{ue}}]_2 - [\boldsymbol{p}_{\textrm{ris}}]_2}{\lVert \boldsymbol{p}_{\textrm{ue}} - \boldsymbol{p}_{\textrm{ris}} \rVert ^2}, \frac{\partial \vartheta}{\partial [\boldsymbol{p}_{\textrm{ue}}]_2} = \frac{[\boldsymbol{p}_{\textrm{ue}}]_1 - [\boldsymbol{p}_{\textrm{ris}}]_1}{\lVert \boldsymbol{p}_{\textrm{ue}} - \boldsymbol{p}_{\textrm{ris}} \rVert ^2}.
\end{equation}

\textit{Scenario 2:} The derivatives are as follows:
\begin{equation}
    \frac{\partial \tau}{\partial [\boldsymbol{p}_{\textrm{ue}}]_1} = \frac{[\boldsymbol{p}_{\textrm{ue}}]_1 - [\boldsymbol{p}_{\textrm{ris}}]_1}{c \lVert \boldsymbol{p}_{\textrm{ue}} - \boldsymbol{p}_{\textrm{ris}} \rVert}, \frac{\partial \tau}{\partial [\boldsymbol{p}_{\textrm{ue}}]_2} = \frac{[\boldsymbol{p}_{\textrm{ue}}]_2 - [\boldsymbol{p}_{\textrm{ris}}]_2}{c \lVert \boldsymbol{p}_{\textrm{ue}} - \boldsymbol{p}_{\textrm{ris}} \rVert},
\end{equation}
\begin{equation} 
    \frac{\partial \vartheta}{\partial [\boldsymbol{p}_{\textrm{ue}}]_1} = - \frac{[\boldsymbol{p}_{\textrm{ue}}]_2 - [\boldsymbol{p}_{\textrm{ris}}]_2}{\lVert \boldsymbol{p}_{\textrm{ue}} - \boldsymbol{p}_{\textrm{ris}} \rVert ^2}, \frac{\partial \vartheta}{\partial [\boldsymbol{p}_{\textrm{ue}}]_2} = \frac{[\boldsymbol{p}_{\textrm{ue}}]_1 - [\boldsymbol{p}_{\textrm{ris}}]_1}{\lVert \boldsymbol{p}_{\textrm{ue}} - \boldsymbol{p}_{\textrm{ris}} \rVert ^2}.
\end{equation}
In both scenarios, it is evident that $\frac{\partial \Re{(\beta_i)}}{\partial \Re{(\beta_i)}} = \frac{\partial \Im{(\beta_i)}}{\partial \Im{(\beta_i)}} = 1.$ All other entries in the Jacobian matrix \(\boldsymbol{\mathcal{J}}_i\) are zero.


\section*{Acknowledgment}


The work of Hui Chen and Henk Wymeersch are supported by the Swedish Research Council under VR grant 2022-03007 and the SNS JU project 6G-DISAC under the EU’s Horizon Europe Research and Innovation Program under Grant Agreement No 101139130.

\ifCLASSOPTIONcaptionsoff
  \newpage
\fi


\bibliographystyle{ieeetr}
\bibliography{reference}

\end{document}